# Observation of a dynamical topological phase transition

**Authors:** N. Fläschner[1,2*], D. Vogel[1*], M. Tarnowski[1,2*], B. S. Rem[1,2], D.-S. Lühmann[1], M. Heyl[4], J. C. Budich[5,6], L. Mathey[1,2,3], K. Sengstock[1,2,3†], C. Weitenberg[1,2]

**Affiliations:**
[1]ILP - Institut für Laserphysik, Universität Hamburg, Luruper Chaussee 149, 22761 Hamburg, Germany
[2]The Hamburg Centre for Ultrafast Imaging, Luruper Chaussee 149, 22761 Hamburg, Germany
[3]ZOQ - Zentrum für Optische Quantentechnologien, Universität Hamburg, Luruper Chaussee 149, 22761 Hamburg, Germany
[4]Physik Department, Technische Universität München, 85747 Garching, Germany
[5]Institute for Quantum Optics and Quantum Information of the Austrian Academy of Sciences, 6020 Innsbruck, Austria
[6]Institute for Theoretical Physics, University of Innsbruck, 6020 Innsbruck, Austria
[*]These authors contributed equally to this work.
[†]Corresponding author. Email: klaus.sengstock@physnet.uni-hamburg.de

**Abstract**: Phase transitions are a fundamental concept in science describing diverse phenomena ranging from, e.g., the freezing of water to Bose-Einstein condensation. While the concept is well-established in equilibrium, similarly fundamental concepts for systems far from equilibrium are just being explored, such as the recently introduced dynamical phase transition (DPT). Here we report on the first observation of a DPT in the dynamics of a fermionic many-body state after a quench between two lattice Hamiltonians. With time-resolved state tomography in a system of ultracold atoms in optical lattices, we obtain full access to the evolution of the wave function. We observe the appearance, movement, and annihilation of vortices in reciprocal space. We identify their number as a dynamical topological order parameter, which suddenly changes its value at the critical times of the DPT. Our observation of a DPT is an important step towards a more comprehensive understanding of non-equilibrium dynamics in general.

**Main Text:** The study of many-body systems far from equilibrium has become a vibrant field of research in physics *(1)*, which has led to the development of several new concepts, but still lacks a systematic understanding compared to the general principles that are available in equilibrium.

Recent experimental efforts have shed light on a variety of fascinating dynamical phenomena, such as prethermalization in closed quantum systems *(2-4)*, many-body localization in disordered interacting systems *(5)*, or light-induced non-equilibrium superconductivity *(6)* to name just a few examples. A promising approach to understand principles of non-equilibrium systems is to extend the concept of phase transitions beyond the conventional equilibrium paradigm. Such an extension was recently proposed by introducing a dynamical counterpart of equilibrium phase transitions *(7)*. While equilibrium phase transitions occur at a critical value of a parameter such as temperature, these dynamical phase transitions (DPTs) occur at critical times after a sudden quench of the system. A formal analogy between temperature and time allows extending elementary concepts of equilibrium statistical physics such as non-analytic behavior, universality, or order to the time-evolution of a quantum state. DPTs have been predicted for different systems including quantum Ising models *(7-14)* and their extensions to non-integrable models *(15-17)*. Very recently, it was suggested that DPTs can also appear in topological lattice Hamiltonians *(18-21)*.

Here, we report on the first experimental observation of a DPT. We study the time evolution of fermionic quantum gases in a hexagonal optical lattice *(22,23)* after a rapid quench from a topologically trivial system into a Haldane-like system *(24)*, which can have either trivial or non-trivial topology. We thus explore the specifically interesting case of a dynamical topological phase transition (DTPT) with possible interesting applications for the characterization of topological matter in general. We employ the recently developed fully momentum-resolved state tomography method *(23)* to identify the dynamical appearance and subsequent disappearance of vortex-anti-vortex pairs in the phase of the wavefunction. We identify their number as a dynamical order parameter, thereby establishing a crucial new ingredient of a DPT. We map out the dynamical phase diagram, which – as a central aspect of DPTs – yields information on the underlying equilibrium phases. It is fascinating to see that the nature of the dynamically appearing defects precisely corresponds to the topological defects that characterize the static system in the form of Dirac points. Since topological phase transitions can occur in non-interacting systems, we have chosen to work with spin-polarized fermions. This allows clearly distinguishing our observations from other many-body phenomena such as Kibble-Zurek mechanism or prethermalization, which also occur after quenches.

Equilibrium phase transitions are signaled by a sudden onset or disappearance of order as a function of a control parameter such as temperature $T$ (Fig. 1A) or pressure. This sudden change of order is associated with a non-analytic behavior of the free energy $F$ at the critical temperature, e.g. a jump of the heat capacity $c_V = -T \mathrm{d}^2 F/\mathrm{d}T^2$. In analogy, a dynamical phase transition can be defined for the time evolution of a quantum many-body system (Fig. 1B), if time $t$ takes the role of an inverse temperature $\beta = 1/(k_B T)$ *(7)*. At critical times, one finds a non-analytic behavior in the dynamical free energy $g(t) = -\ln(|G(t)|^2)$, which is defined in analogy to the free energy $F(\beta) = -(1/\beta)\ln(Z(\beta))$ of the equilibrium case. Here $G(t) = \langle \psi_0 | \exp(-iHt/\hbar) | \psi_0 \rangle$ is the Loschmidt amplitude, which resembles the equilibrium boundary partition function $Z(\beta) = \langle \psi | \exp(-\beta H) | \psi \rangle$ and describes the evolution of a quantum state $|\psi_0\rangle$ under the Hamiltonian $H$. To push this analogy further, we show within this work that one can also define a dynamical topological order parameter, which changes its value at the critical times.

In our experiment, we use a quench between a static and a Floquet Hamiltonian to initialize the dynamics. We start with a band insulating state in the lowest band of a hexagonal lattice with a large offset between the A and B sites (Fig. 2A). We then quench into the final Floquet Hamiltonian, which is engineered by means of resonant circular lattice shaking *(23,25)*. The circular shaking breaks time-reversal symmetry, such that the final Haldane-like Hamiltonian can have bands with non-trivial Chern numbers *(26)*. In a two-band model, the Hamiltonian and the time-evolved state can be visualized on a Bloch sphere for each quasimomentum (Fig. 2B). If we choose the poles of the Bloch sphere as the eigenstates of the initial Hamiltonian, the initial states point to the north pole. After the quench, the final Hamiltonian points to different directions for each quasimomentum leading to a momentum-dependent dynamics of the state on the Bloch sphere (Fig. 2B). We measure this dynamics via a momentum and time-resolved state tomography *(23)* after the stroboscopic evolution times of the Floquet Hamiltonian and visualize the ensuing many-body state on the Bloch sphere (Fig. 2C) *(27)*.

When the evolved many-body state has become orthogonal to the initial state, the Loschmidt amplitude $G(t)$ has decayed to zero. In analogy to the equilibrium case, this is called a Fisher zero *(7)* and leads to the non-analytic behavior of the dynamical free energy $g(t)$, which is associated with the DPT. On the Bloch sphere, this means that the many-body state reaches the south pole for one quasimomentum. Because neighboring quasimomenta do not reach the south pole at the same time and the geometry of the eigenstates is a smooth function of momentum, this has to give rise to a vortex in the azimuthal phase profile $\varphi_k(t)$. These vortices are a direct manifestation of the Fisher zeros of the underlying Loschmidt amplitude. Their existence can therefore serve as a dynamical order parameter *(27)*.

As a key result of this work, we observe these dynamical vortices in the azimuthal phase profile in our time-resolved data (Fig. 3A-C). The dynamical vortices appear and annihilate as vortex-anti-vortex pairs and therefore their movement traces out a closed contour in momentum space (Fig. 3D). Single vortices cannot appear in our time evolution, because the Chern number of a state cannot change under unitary evolution *(28,29)* and is therefore fixed to the topology of our initial bands. In particular, a dynamically defined Chern number would not constitute a suitable order parameter. In addition to the dynamical vortices, the phase profile $\varphi_{\boldsymbol{k}}(t)$ also has the expected static vortices at the position of the Dirac points of the final Hamiltonian.

Our system illustrates the celebrated robustness of topological order, because the appearance of dynamical vortices is not hampered by a slight breaking of the threefold symmetry of the states. This asymmetry stems from the initial phase of the lattice shaking *(30)* and from experimental imperfections such as tiny beam imbalances or weak multi-photon coupling to higher bands *(31)*.

Finally, we map out the dynamical phase diagram by repeating our measurements for different lattice depths and thereby realizing different detunings $\delta = \Delta_{AB} - \nu_{sh}$ between the AB-sublattice offset $\Delta_{AB}$ and the shaking frequency $\nu_{sh}$ (Fig. 4A,B). This detuning effectively determines the energy distance between the two Floquet bands. For suitable detunings, this can lead to an opening and closing of the Dirac points such that the Chern number of the Floquet bands can be changed to non-trivial values (Fig. 4C). We can interpret our results as a 2D dynamical phase diagram spanned by detuning and time, in which dynamical order appears in a distinct area. The critical times of $t_c \cong 0.5$ ms and 1.0 ms nicely reflect the scale of the band energy distances in the Floquet Hamiltonian.

We compare our results to the topological phase diagram of the final Floquet Hamiltonian, which is a Haldane-like model *(24)* with non-trivial Chern numbers for suitable detunings (Fig. 4C). We can thus address the important question of the intimate relation between DPTs and the underlying phase diagram *(7,10)*. Starting from a lattice with trivial topology, the Chern number changes across the quench for a narrow region in detuning of 80 Hz width. In this case, the appearance of a DPT is ensured *(18)*. We observe the DPT for a broader range of detunings of 280 Hz width, which rather corresponds to the AB-tunneling element of the initial bands of

$J_{AB} \cong 260$ Hz. This observation demonstrates that the DPT is not only sensitive to a change of the topology across the quench, but also to a strong change of the geometry of the eigenstates, and can therefore serve as a precursor for the topologically non-trivial region *(27)*. Our results illustrate the intimate connection between the ground state phase diagram and the appearance of dynamical order in the non-equilibrium dynamics after a quench within the phase diagram. This opens up an interesting approach for studying ground state phase diagrams in cases where it is experimentally challenging to reach the ground state.

Our measurements constitute the first realization of a DPT and not only demonstrate that the concept applies for very general intriguing studies of out-of-equilibrium physics but also pave the way for a deeper understanding of quantum phases via rapid quenches across phase boundaries. An obvious extension of our studies involves interacting quantum gases. Such systems would be ideally suited to study the interplay and differences between various concepts such as the Kibble-Zurek mechanism, prethermalization and DPTs, which all describe the dynamical evolution after a quench. In condensed matter physics, DPTs could constitute a new route towards the exploration of phases at extreme conditions such as highest pressures or magnetic fields, which usually are not accessible in a continuous way. While it is often experimentally challenging to reach the ground state of relevant systems, the observation of dynamical order in highly-excited states might be an alternative approach to the exploration of the underlying ground state phase diagram, in cases where a thorough connection between static and dynamical order can be established *(10)*. This might open a promising route to study complex phase diagrams such as the Fermi-Hubbard model or interacting topological phases.

**Acknowledgments:** We acknowledge financial support by the excellence cluster "The Hamburg Centre for Ultrafast Imaging - Structure, Dynamics and Control of Matter at the Atomic Scale" and the GrK 1355 of the Deutsche Forschungsgemeinschaft. BSR acknowledges financial support from the European Commission (Marie Curie Fellowship), MH from the Deutsche Akademie der Naturforscher Leopoldina (Grant No. LPDR 2015-01), and JCB from the ERC synergy grant UQUAM.

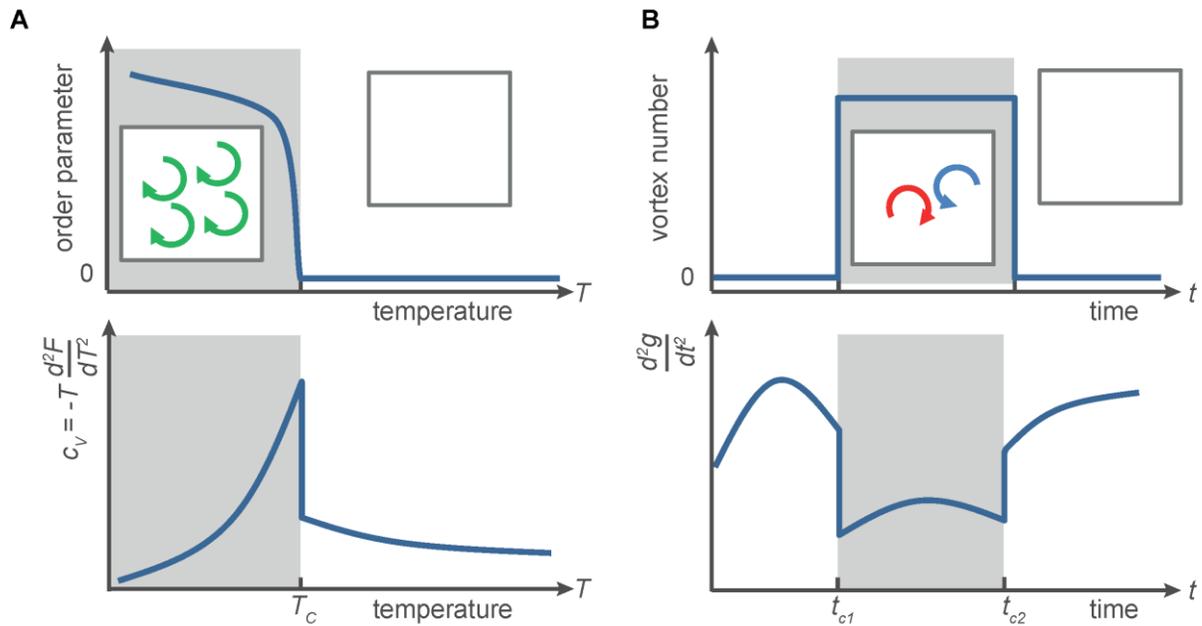

**Fig. 1. Schematic comparison between equilibrium and dynamical phase transitions.** (**A**) In a thermal equilibrium phase transition, the order parameter e.g. vanishes at the critical temperature, where a derivative of the free energy $F(T)$ has a jump. (**B**) In analogy, in a dynamical topological phase transition, topological order appears and vanishes at critical times in the quantum dynamics of a many-body system, where a derivative of the dynamical free energy $g(t)$ has a jump. The gray shading indicates the parameter regime with static or dynamical order. The insets schematically illustrate possible order parameters, e.g. with vorticity.

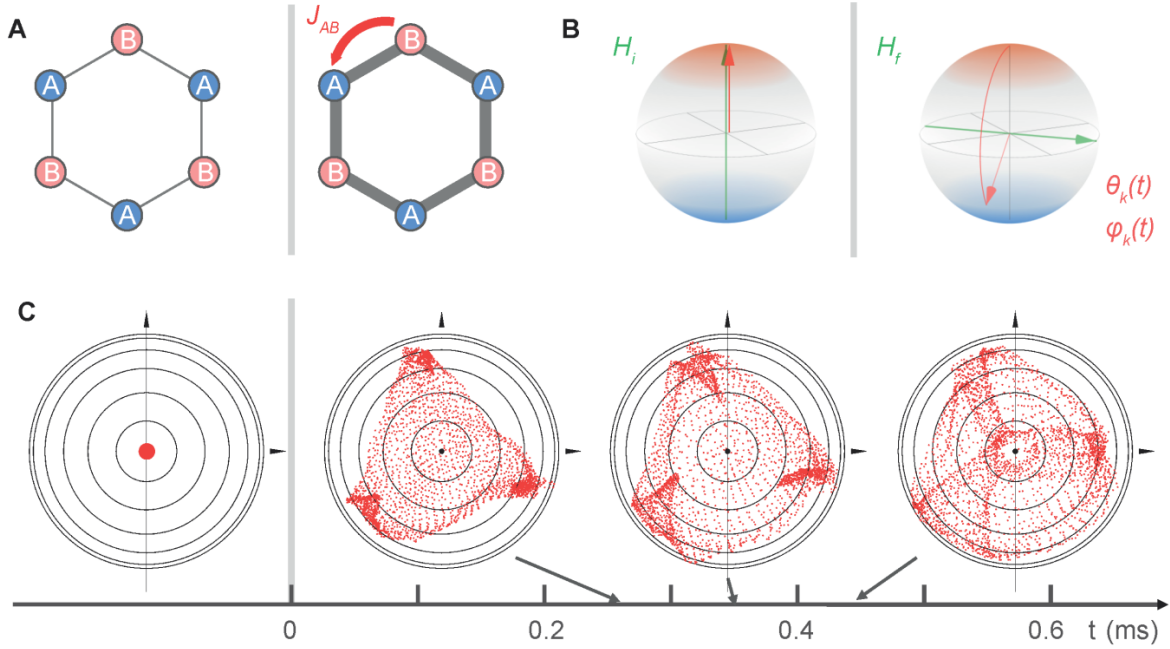

**Fig. 2. Quantum dynamics after the quench between two lattice Hamiltonians.** (**A**) Our initial hexagonal lattice has a large offset between the A and B sublattices such that tunneling is essentially suppressed and the atoms are localized on the B sites. At $t = 0$, we suddenly switch on the tunneling and initiate the dynamics. (**B**) At each quasimomentum $\mathbf{k}$, the state can be described on a Bloch sphere, whose poles are given by the lower and upper band of the initial Hamiltonian $H_i$, and the initial state points to the north pole for all quasimomenta. The final Hamiltonian $H_f$ points to different directions leading to a $\mathbf{k}$-dependent rotation dynamics on the Bloch sphere with the instantaneous azimuthal and polar angles $\varphi_\mathbf{k}(t)$ and $\theta_\mathbf{k}(t)$ of the time-evolved state. (**C**) Using our state tomography technique *(23)*, we measure the time-evolved state for each quasimomentum in the first Brillouin zone and reconstruct the instantaneous many-body state. This data is represented on the Bloch sphere (viewed from the north pole) for different evolution times after the quench. The data shows that the individual quasimomentum states spread around the Bloch sphere and reach the south pole.

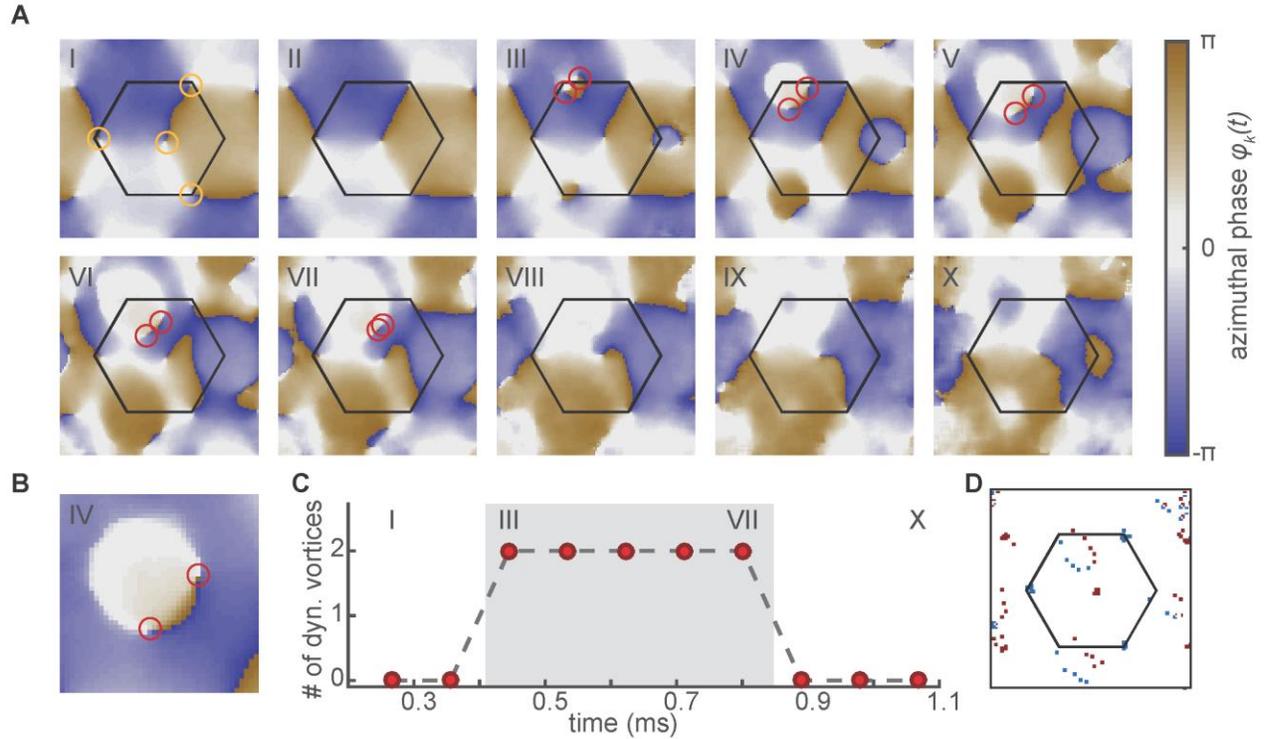

**Fig. 3. Observation of a dynamical topological phase transition.** (**A**) Experimental phase profiles $\varphi_{\boldsymbol{k}}(t)$ for different stroboscopic evolution times (multiples of 89 μs starting with 267 μs; labeled by Roman numerals). The hexagon marks the first Brillouin zone. Static vortices (marked by orange circles in (I)) are imprinted from the final Hamiltonian *(23)* and remain fixed for all times. Additionally, dynamical vortices appear, move and disappear during the time evolution (marked by red circles). (**B**) Zoom into the phase profile at time step (IV) demonstrating also our high momentum resolution. (**C**) The number of dynamical vortices in the first Brillouin zone serves as a dynamical order parameter, which changes at the critical times (grey shading indicates the presence of dynamical order). (**D**) Position of the vortices summed over all observed evolution times (blue and red dots for clockwise and counter-clockwise phase winding, respectively). The dynamical vortices are created and annihilated as vortex-anti-vortex pairs and they move on a closed contour in momentum space.

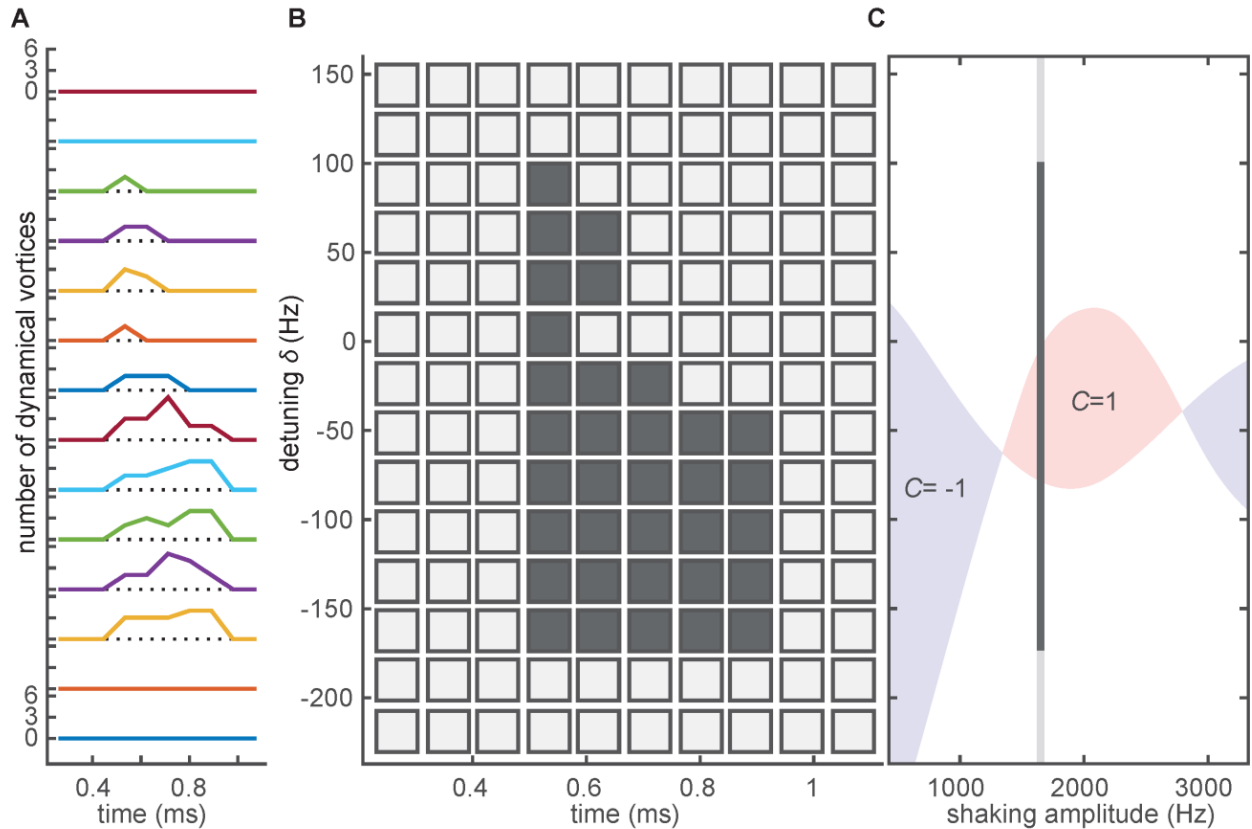

**Fig. 4. Dynamical phase diagram and its relation to the equilibrium phase diagram.** (**A**) Number of dynamical vortices as a function of evolution time for different detunings $\delta$ between the AB-offset and the shaking frequency *(27)* (color coded; the detuning values are the same as those of the corresponding line in (**B**)). (**B**) Dynamical phase diagram spanned by the detuning and the time obtained from the data in (**A**). Filled boxes refer to the presence of dynamical order. (**C**) The topological phase diagram of the underlying final Floquet Hamiltonian as a function of detuning and shaking amplitude features a non-trivial Chern number for suitable parameters. The vertical bar denotes the path of endpoints of the fast quench that was traced out in the experiment (shaking amplitude 1.65 kHz) and the dark color indicates the region with DPT.

**Supplementary materials:**
Supplementary Text
Figs. S1-S5
References *(32)* to *(35)*

# Supplementary Materials

## Content
S.1 Optical lattice setup
S.2 Data acquisition and analysis
S.3 Calibration of the lattice depth
S.4 Analysis of vorticity
S.5 Amplitude data
S.6 Relation between dynamical and equilibrium phase diagram
S.7 Relation to Loschmidt amplitude and Fisher zeros
S.8 Calculation of the Floquet phase diagram

## S.1 Optical lattice setup

For the experiments presented in the main text, we employ a hexagonal optical lattice with tunable energy offset $\Delta_{AB}$ between the A- and B-sublattice on-site energies. The lattice is formed by three running interfering laser beams (Fig. S1). For more details on the lattice setup, see Ref. *(22,23)*.

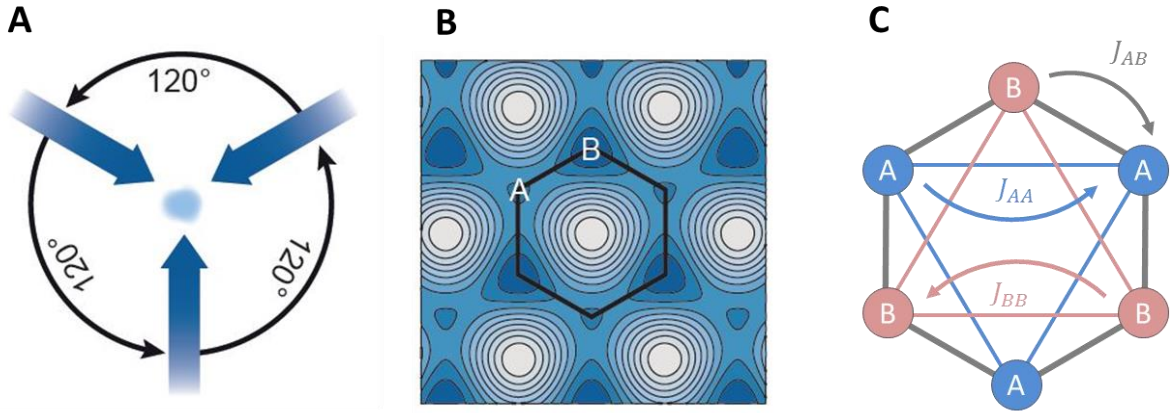

**Fig. S1: Setup of the optical lattice**. (A) The lattice is formed by three interfering laser beams. (B) Using tunable polarizations, the resulting potential can have an energy offset between the A and B sublattice. (C) Hopping amplitudes $J_{AA}, J_{BB}, J_{AB}$ in the tight-binding description of the lattice.

The two lowest bands of this lattice can be well described within a tight-binding model with the eigenbasis being the Bloch states restricted to the A- and B-sublattice sites *(32)*. In this basis, the quasimomentum-dependent Hamiltionian reads

$$\hat{\mathbf{H}}(\mathbf{k}) = \begin{pmatrix} \dfrac{\Delta_{AB}}{2} + \sum_i 2 J_{AA} \cos(\mathbf{k} \cdot \mathbf{a}_i) & \sum_i J_{AB} \exp(-i\mathbf{k} \cdot \mathbf{d}_i) \\ \sum_i J_{AB} \exp(i\mathbf{k} \cdot \mathbf{d}_i) & \dfrac{-\Delta_{AB}}{2} + \sum_i 2 J_{BB} \cos(\mathbf{k} \cdot \mathbf{a}_i) \end{pmatrix}$$

with the next-nearest-neighbor hopping amplitudes $J_{AA}$ and $J_{BB}$ on the diagonal and the nearest-neighbor hopping amplitude $J_{AB}$ on the off-diagonal element, which couples the A- and B-sites (see Fig. S1).

The Bravais vectors connecting neighboring A(B)-sites with each other are, for our purely hexagonal lattice structure, $a_1 = -4\pi/(3k_L)(0\;\;1\;\;0)$, $a_2 = -4\pi/(3k_L)(\sqrt{3}/2\;\;1/2\;\;0)$, and $a_3 = a_2 - a_1$. The $d_i$ connect A- with neighboring B-sites $d_1 = 1/3(a_1 + a_2)$, $d_2 = 1/3(-2a_1 + a_2)$, and $d_3 = 1/3(a_1 - 2a_2)$. The lattice vector $k_L$ is given by the wavelength of the laser forming the lattice $k_L = 2\pi/1064\mathrm{nm}$.

All tight-binding parameters of the lattice can be expressed as a function of the single parameter $\Delta_{AB}$, once the polarizations of the lattice beams, which determine the geometry, are fixed *(23)*. We therefore use $\Delta_{AB}$ as a measure for the lattice depth throughout the manuscript. For our polarization setting and for the parameter regime of $\Delta_{AB}$ presented in the main text, we determine the relations from a fit to exact band structure calculations:

$$J_{AB}/E_R \approx 0.221 \cdot (\Delta_{AB}/E_R)^{-1.4},\; J_{AA}/E_R \approx 0.0529 \cdot (\Delta_{AB}/E_R)^{-2.05},\text{ and } J_{BB} \approx 0.$$

Here $E_R$=4,410 Hz is the recoil frequency. All data was taken in a configuration with large $\Delta_{AB}$ ranging from 10,750 Hz to 11,400 Hz, which leads to tunneling amplitudes $J_{AB}$ ranging from 280 Hz to 258 Hz and $J_{AA}$ ranging from 38 Hz to 33 Hz, such that $\Delta_{AB}$ is the dominating energy scale and the initial bands are nearly flat.

Since the position of the lattice is determined by the relative phases of the three running laser beams forming the lattice, we can move the lattice in real space by frequency modulating the laser beams using acousto-optical modulators (AOMs). In order to realize circular lattice shaking as used for all experiments presented in the main text, two of the three AOMs are driven by computer programmable digital frequency sources with a time-dependent frequency $v_{1,2}(t) = 110\mathrm{MHz} + 2A_s[\pm\cos(2\pi v_s t) + \sqrt{3}\sin(2\pi v_s t)]$, where $A_s$ and $v_s$ are referred to as shaking amplitude and frequency, respectively. The shaking frequency is $v_s$=11,236 Hz throughout the manuscript and the shaking amplitude is $A_s$=2 kHz in Fig. 2 and 3 and 1.65 kHz in Fig. 4. With this setup we have full control of the shaking phase *(30)*.

### S.2 Data acquisition and analysis

Our experiments start with a spin polarized cloud of $1\times10^5$ fermionic $^{40}$K atoms in a non-interacting band insulator in the lowest band of the two-dimensional initial lattice. In the transverse direction, the potential is harmonic, thus forming a lattice of tubes. The external confinement from the shape of the lattice beams and additional dipole trap beams leads to trapping frequencies of $v_{x,y,z} = (83(4), 108(5), 93(4))\mathrm{Hz}$.

The full state tomography method (FST) is based on the ideas of Ref. *(33,34)* and was demonstrated and described in detail in Ref. *(23)*. The main idea is to project back onto the initial bands and to observe the ensuing quench dynamics on the Bloch sphere, which translates into a quasimomentum-dependent density distribution after a time-of-flight expansion due to the interference of the bands. By a pixelwise evaluation of the amplitude and phase of this oscillation, we obtain the momentum-resolved polar and azimuthal phase profiles of the original many-body state.

To resolve this quench dynamics, we take 32 images at varying times $t_{FST}$ after the quench onto the initial bands with a step size of $5.5\mu s$. In the data analysis, we apply a 5x5 pixel Gaussian filter of width 5 pixel on the atomic density of each time step, but no temporal filtering. With a length of the reciprocal lattice vector of 58(2)px, we have more than 2800 pixels in the first Brillouin zone and therefore a very high resolution in momentum space. For every pixel, we fit a damped oscillation to the density of the form

$$n(k,t_{FST}) = W_k[1 + \alpha_k \exp(-\gamma_k t_{FST})\sin(2\pi v_0 t_{FST} + \varphi_k)]$$

with the envelope $W_k$, the oscillation amplitude $\alpha_k$, the damping $\gamma_k$, the oscillation frequency $v_0$ and the phase $\varphi_k$. In order to minimize the number of fit parameters, we fix the frequency in the fit to $v_0 = 11.364\text{kHz}$, which is the third Fourier component for these time steps. We initialize the phases by the values from a fast Fourier transform estimator and the amplitudes from the difference of the minimal and maximal values.

### S.3 Calibration of the lattice depth

In order to map out the very narrow region of the DPT in the phase diagram, we need a much more precise lattice depth calibration than usually required and we achieve this in two steps. As a first step, we perform lattice amplitude modulation in each of the three different 1D-lattices that are generated when only two laser beams interfere, while the third laser beam is detuned with respect to the other ones. This calibration yields a systematic uncertainty of the lattice depth $V_{1D}$ of about 1% due to the broadening of the frequency response due to the harmonic trap *(35)* and is only used to ensure relative beam balance between the three laser beams.

In order to determine the lattice depth in the 2D-lattice formed by the three laser beams with larger precision, we use the fact that the frequency of the observed momentum density oscillation after the projection back onto the initial bands corresponds to their energy distance. Due to the residual curvature of the bands, this frequency is momentum-dependent. We accordingly repeat the fit discussed in Section S.2 with the momentum-dependent frequency $v_k$ as a free parameter.

The amplitude is necessarily zero at the phase vortices and small around the vortices. In order to enhance the reliability of the lattice depth estimation, we thus only use the frequencies obtained for momenta well away from the vortices. For this range of momenta (covering approximately 50% of the Brillouin zone), the mean frequency $v_m$ approximately corresponds to the AB-offset $\Delta_{AB} \approx 0.9964 v_m$. In order to avoid imprecise fit results due to the appearance and movement of the dynamical vortices, we only use the data obtained for lattice depths for which the dynamical phase transition does not occur ($\Delta_{AB}$ =10,750 Hz to 11,100 Hz, data not completely shown in Fig. 4 of the main text). Furthermore, we only use the data of the four shortest stroboscopic evolution time steps (*t*=89, 178, 267, 356 µs) for which the damping is small and the influence of the harmonic trap is negligible.

From the slope $S$ of a linear fit ($\Delta_{AB} = S \cdot V_{1D}$) to the data obtained for ten differently set 1D-lattice depths $V_{1D}$ (Fig. S2), we infer the lattice depth quantified by $\Delta_{AB}$ as stated in the main text. This slope changes by $3 \cdot 10^{-3}$ when taking more or less evolution time steps into account, which gives an estimate of the systematic error of the calibration.

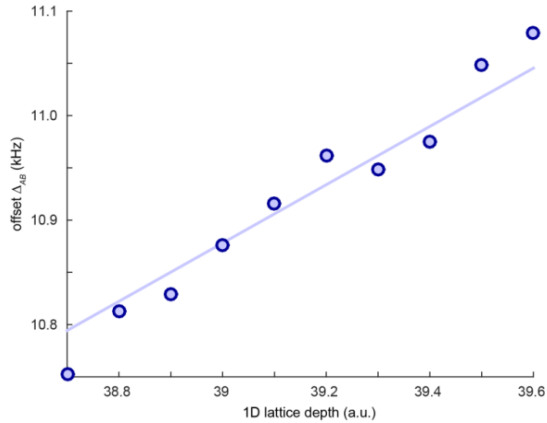

**Fig. S2: Lattice depth calibration**. The mean frequency of a momentum-dependent fit for various 1D lattice depths (blue datapoints) yields the AB-offset $\Delta_{AB}$. A linear fit (blue line) to the data is used to calibrate the lattice depth.

### S.4 Analysis of vorticity

To identify the vortices in our phase profiles $\varphi_k(t_n)$, we calculate the vorticity for each evolution time $t_n$. It is determined as the curl of the gradient of $\varphi_k(t_n)$. Usually this quantity is zero, but for phase profiles, which are defined modulo $2\pi$, it obtains finite values at the singular points, i.e. at the vortices. It is therefore a convenient measure of the local vorticity. The derivatives are evaluated as finite differences on the natural pixel grid of our images, and the gradient is furthermore symmetrized by averaging of the left and right difference. The position of the vortices is determined to the precision of one pixel and the routine yields a finite vorticity on four neighboring pixels at a given vortex.

The local nature of this quantity makes it suitable to well resolve the annihilation of the vortex pairs, but also makes it susceptible to experimental noise in the phase profiles. Noise can produce many closely spaced vortex-anti-vortex pairs. However, due to our excellent data quality, this happens only far outside the first Brillouin zone (FBZ) and for long evolution times. It is therefore not necessary to apply further spatial filtering.

Fig. 3D is obtained by adding up the vorticity images of the time steps $t_n$ with $n=3...12$. The resulting trace illustrates the movement of the vortices along their contours. The colorbar is truncated to the value of a single vortex in order to avoid large values from the static vortices, where the signal adds up. The error for a wrong count of the vortex number within the FBZ is therefore very small and certainly smaller than the symbol size in Fig. 3C.

For Fig. 4A we count the dynamical vortices within the FBZ for each evolution time using the vorticity and cross checking with the original phase fields. The static vortices can always be clearly identified and are not counted here. We see that the dynamical vortices always appear and annihilate in pairs. Therefore one might expect that the dynamical vortex numbers should always be even and should not change value within the dynamically ordered phase. However, our dynamical vortex numbers do change and can have odd values. This happens, because the threefold symmetry of the states is broken due to the initial shaking phase and due to experimental imperfections. The three vortex pairs do not always appear in the same time step and single vortices can move out of FBZ, where they are not counted, while the partner remains within the FBZ. One could easily enforce the even number of dynamical vortices by counting the pairs in the vicinity of the FBZ, but we choose to count only within the FBZ in order to have a clear and reproducible protocol.

## S.5 Amplitude data

Our state tomography *(23)* yields the many-body quantum state via the phase profile $\varphi_k$ and the amplitude profile $\sin(\theta_k)$, which correspond to the azimuthal and polar angle in a Bloch sphere description. In the main text, we restrict the discussion to the phase profile $\varphi_k$; in Fig. S3 we show the amplitude profile corresponding to Fig. 3. As expected, the phase vortices go along with an amplitude of zero at the vortex core, i.e. $\theta_k = 0, \pi$.

Note that our state tomography can a priori not distinguish between the northern and the southern hemisphere of the Bloch sphere, because we measure the amplitude $\sin(\theta_k)$ instead of the polar angle $\theta_k$ itself. In practice, however, one can always retain the information about the hemisphere by following the time evolution. In Fig. S4, the vortices at the north and south poles both appear as zero amplitudes.

In Fig. S3 one recognizes a quasimomentum-dependent damping in the system. E.g. in image V, the region of zero amplitude at the static vortices correspond to the north pole, while the region of zero amplitude of the dynamical vortices corresponds to the south pole. In between, the state has to cross the equator and yield an amplitude of 1. However, it only goes up to 0.7, while the amplitude does have values close to 1 at the edge of the Brillouin zone. This momentum-dependent damping might have its origin in a combination of band decay and the external trapping potential, which can induce dynamics in incompletely filled bands *(35)*. To compensate for this damping, the amplitudes in Fig. 2C have been scaled by 14%, 21%, and 28%, such that the maximal amplitudes reach the equator. While the damping is clearly visible in the amplitude data, it cannot destroy the phase vortices and therefore doesn't affect the DPT.

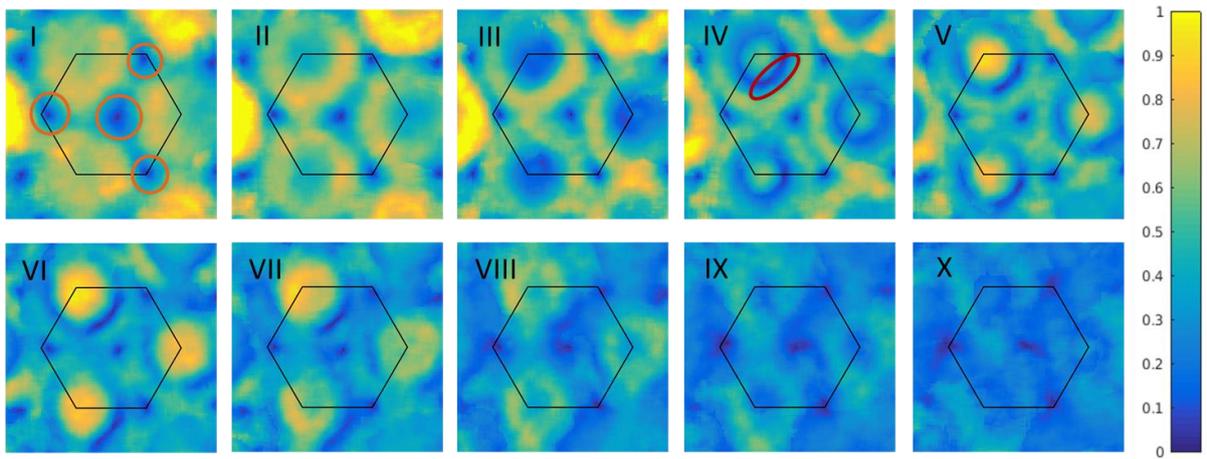

**Fig. S3: Amplitude data**. Amplitude data for the quantum dynamics belonging to the phase data in Fig. 3. The roman numbers give the same time steps as in Fig. 3. The orange circles mark the static vortices (visible as points of zero amplitude). The red ellipse marks the ring of small amplitude, where the vortex pair appears.

## S.6 Relation between dynamical and equilibrium phase diagram

In Fig. 4, we compare the region with DPT in the dynamical phase diagram to the region of Chern number one in the static phase diagram and discuss the relation between them. A change in topology across the quench will always lead to a DPT in the dynamics, as proven in Ref. *(18)* (the absolute value of the Chern number has to change across the quench). Our measurements show that the inverse conclusion does not hold. Instead, a strong change of the geometry of the eigenstates across the quench, as it appears close to the regions of nontrivial Chern number, is sufficient to induce a DPT. A similar behavior was found for the 1D case *(18,20)*.

To understand this, it is useful to distinguish two bases, i.e. two sets of quasimomentum-dependent Bloch spheres for the description of the many-body state. In the main text, we define a Bloch sphere spanned by the eigenstates of the initial lattice. We refer to this as the relative Bloch sphere between initial and final Hamiltonian in the following. Our state tomography naturally measures in this basis, because it involves a projection back onto the initial nearly flat bands. In this basis the quasimomentum states are all perfectly initialized on the north pole.

There also is another natural basis for the description of a two-band model, which is the basis of completely flat bands with $J_{AB} = 0$, which are completely localized on the A and B sublattices, respectively (which is the basis of the tight-binding model of Section S.1). We refer to this basis as the absolute Bloch sphere. The topological properties of the final Hamiltonian can best be visualized in this basis: a Chern number one means that the quasimomentum states within the first Brillouin zone wrap once around the Bloch sphere. Our final Floquet Hamiltonian can have Chern number zero or one depending on the parameters of the lattice shaking. While Chern number one means that the states completely cover the absolute Bloch sphere, for Chern number zero the states remain only on the northern hemisphere. The southern hemisphere is only reached for Chern number one. This is because the Chern number can only change from zero to one, when the bands touch and at this transition, the south pole is the first point on the Southern hemisphere to be reached. Note that these statements are specific to our Floquet Hamiltonian obtained by resonant lattice shaking.

Let us consider the requirements for dynamical topological order to appear in the relative Bloch sphere: the many-body state is initialized as pointing to the north pole for all quasimomenta. The time-evolved state becomes orthogonal to the initial state, when it reaches the south pole for one quasimomentum. This shows as a vortex in the phase profile, because the neighboring quasimomentum states reach the neighboring points on the relative Bloch sphere. Obviously, this requires that the final Hamiltonian reaches the equator and extends into both hemispheres.

Now the distinction between the two Bloch spheres becomes important: if we transform the final Hamiltonian into the relative Bloch sphere, it can reach the Southern hemisphere without covering it completely. This situation leads to the DPT without change of Chern number across the quench. One can also picture this situation in the absolute Bloch sphere: because the initial bands are not completely flat, they cover a finite area around the north pole and the states are initialized correspondingly. Now a final Hamiltonian close to the equator, but still on the northern hemisphere, would be sufficient to bring these states to the opposite side of the Bloch sphere close to the south pole during the time evolution.

In the particular case of a quench starting from a Hamiltonian with completely flat bands, in which the two bases are identical, the DPT can only appear when the final Hamiltonian has non-trivial

Chern number. But for our Floquet system, realized by dressing the initial bands, this would lead to an infinitely small region of Chern number one. Instead, our setup allows studying a variety of quench scenarios between two Hamiltonians with different geometry or different topology.

## S.7 Relation to Loschmidt amplitude and Fisher zeros

There is an intimate relation between the non-analytic behavior of the dynamical free energy $g(t) = -\ln(G(t))$ at the critical times and our dynamical topological order parameter, because both are related to the Fisher zeros of the Loschmidt amplitude $G(t) = \langle \psi_0 | exp(-iHt/\hbar) | \psi_0 \rangle$. The Fisher zeros manifest themselves as dynamical vortices in the phase profile, which are introduced as dynamical order parameter. While in 1D, the Fisher zeros are only observable at the critical times, in our 2D topological system they remain observable during the dynamically ordered phase *(18)*. Instead their density diverges at the critical times *(18)*. This is why we can introduce the Fisher zeros themselves as dynamical order parameter. As a consequence, the order parameter alone gives insight into the relevant behavior of the Loschmidt amplitude and is sufficient for the clear interpretation as a DPT.

In principle, also the Loschmidt amplitude could be reconstructed from our full state tomography, but the stroboscopic time steps of our Floquet system would not allow observing a non-analytic behavior as a function of time. Instead, we show a calculation, in which the time evolution in the effective Floquet Hamiltonian can also be evaluated at arbitrary times. We start from a calculated final Hamiltonian for typical experimental parameters ($\Delta_{AB} = 11.190\text{kHz}$ $\nu_s = 11.236\text{kHz}$, $A_s = 2\text{kHz}$) and calculate the non-analytical part of the dynamical free energy according to Refs. *(7,18)*. We integrate over only one third of the Brillouin zone, because the Floquet Hamiltonian has a slightly broken symmetry due to the initial shaking phase.

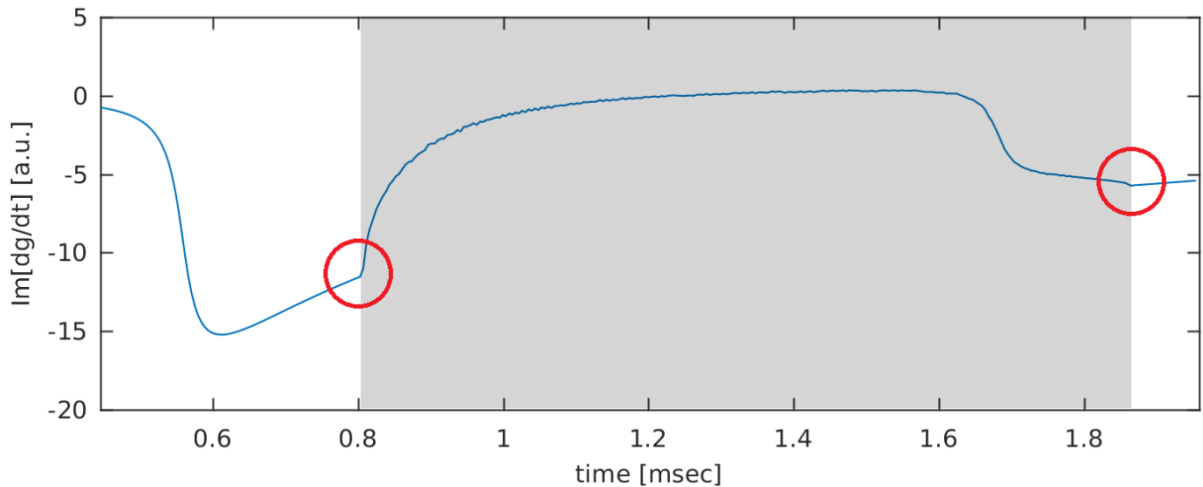

**Fig. S4: Non-analytic behavior of the dynamical free energy.** The plot shows the imaginary part of the first derivative of the dynamical free energy $g(t)$ calculated for a typical final Hamiltonian. One clearly recognizes the expected kinks at the critical times (indicated by red circles), where the dynamical order (grey shading) appears and disappears.

In Fig. S4, we plot the imaginary part of the first derivative of this dynamical free energy. From the plot we can identify the nature of the non-analytic behavior of the dynamical free energy at the critical times: the kink occurs in the first derivative, leading to a jump in the second derivative, as sketched in Fig. 1B. This behavior is also expected from other calculations of 2D topological systems *(18)*. The calculation confirms that indeed our system is described by a DPT in the sense originally introduced in Ref. *(7)* based on the non-analytic behavior of the dynamical free energy.

## S.8 Calculation of the Floquet phase diagram

In order to calculate the Floquet phase diagram as shown in Fig. 4C of the main text, we numerically calculate the Floquet Hamiltonian as the time-averaged Hamiltonian $H_{\text{eff}}(k)$ over one period of the driving $T$,

$$H_{\text{eff}} = \frac{i\hbar}{T} \log[\exp(-\frac{i}{\hbar}\int_0^T H(k,t)\mathrm{d}t)]$$

where $H(k,t)$ is the instantaneous Hamiltonian. The integral is approximated by a time-ordered product of piecewise constant Hamiltonians $\exp(-i/\hbar\, H(k,t_0)\Delta t)$. The matrix exponentials and logarithms can be conveniently evaluated analytically since the Hamiltonian is a 2x2 matrix for each quasimomentum $k$ (see Section S.1) and can thus be expressed in terms of Pauli matrices. The effect of the driving can in an interaction picture be incorporated as $H(k,t) = H(k - \Delta k(t))$ such that the acceleration in real-space translates into a shift in quasimomentum. We perform the calculation on a discrete grid in momentum space with a reciprocal lattice vector corresponding to 56 grid points.

We identify the phase boundaries of the topological phase transition as the parameters at which the Floquet bands touch. We confirm the change in topology by calculating the Chern number $C$ of the resulting Floquet bands. We sum up the Berry curvature $B(k)$ within the first Brillouin zone $C = \frac{1}{2\pi}\sum_k B(k)$. For the calculation of the Berry curvature, it is convenient to express the Floquet Hamiltonian in terms of the Pauli matrices $\sigma_i$ and the identity matrix $I$, $H_{\text{eff}}(k) = h_0(k)I + \sum_{i=1}^{3} h_i(k)\sigma_i$. Then, the Berry curvature is given by

$$B(k) = \frac{1}{2}\hat{h}(k)\cdot\left(\partial_{k_y}\hat{h}(k)\times\partial_{k_x}\hat{h}(k)\right)$$

with $\pm\hat{h}(k)$ being the eigenstates of the Floquet Hamiltonian, $\hat{h}(k) = (h_1(k)\ \ h_2(k)\ \ h_3(k))/\sqrt{h_1^2(k)+h_2^2(k)+h_3^2(k)}$ *(34)*. The eigenstates can be accordingly expressed in terms of $\varphi_k$ and $\theta_k$, $\hat{h}(k) = (\sin(\theta_k)\cos(\varphi_k)\ \ \sin(\theta_k)\sin(\varphi_k)\ \ \cos(\theta_k))$, which correspond to the azimuthal phase and polar angle in the absolute Bloch sphere picture.

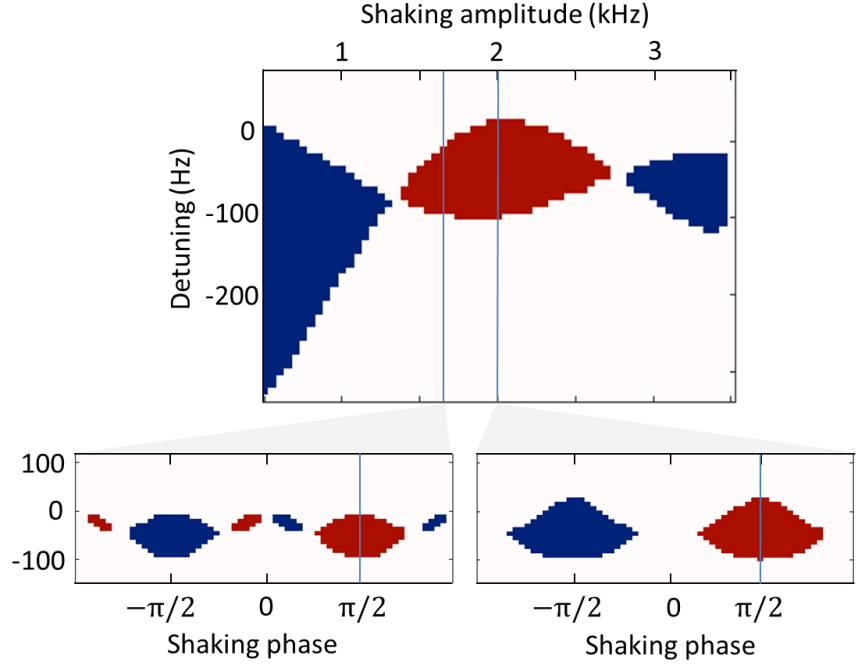

**Fig. S5: Floquet phase diagram.** Static phase diagram of the Haldane-like Floquet Hamiltonian featuring regions with Chern number 0 (gray area), 1 (red area) and -1 (blue area).

We call our Floquet phase diagram Haldane-like, because it describes bands which can have non-zero Chern number in the absence of a net magnetic flux like the famous Haldane model *(24)* (Fig. S5). The detuning between the shaking frequency and the AB-offset of the initial bands is analogous to the AB-offset of the Floquet bands in the Haldane model. When this detuning, which quantifies the breaking of the inversion symmetry, becomes too strong the nontrivial topology is destroyed. The analogy becomes even more illustrative, if one plots the phase diagram as a function of detuning and shaking phase for a fixed shaking amplitude (insets for $A_s = 1.65\text{kHz}$ and $A_s = 2\text{kHz}$). The shaking phase is analogous to the phase of the complex tunneling element in the Haldane model and quantifies the breaking of time-reversal symmetry. For $A_s = 2\text{kHz}$, the phase diagram shows the characteristic two lobes around the maximal time-reversal symmetry breaking at a phase of $\pm\pi/2$ well-known from the Haldane model *(24)*. For $A_s = 1.65\text{kHz}$, additional smaller lobes appear. In the experiment, we work at maximal width of the topologically non-trivial regime realized by circular lattice shaking (blue line).